\let\includefigures=\iffalse 
%
\batchmode
  \font\blackboard=msbm10
\errorstopmode
\newif\ifamsf\amsftrue
\ifx\blackboard\nullfont
  \amsffalse
\fi
\newfam\black
%

%
%
\input harvmac.tex
\includefigures
\message{If you do not have epsf.tex (to include figures),}
\message{change the option at the top of the tex file.}
\input epsf
\def\figin{\epsfcheck\figin}\def\figins{\epsfcheck\figins}
\def\epsfcheck{\ifx\epsfbox\UnDeFiNeD
\message{(NO epsf.tex, FIGURES WILL BE IGNORED)}
\gdef\figin##1{\vskip2in}\gdef\figins##1{\hskip.5in}
\else\message{(FIGURES WILL BE INCLUDED)}%
\gdef\figin##1{##1}\gdef\figins##1{##1}\fi}
\def\DefWarn#1{}
\def\figinsert{\goodbreak\midinsert}
\def\ifig#1#2#3{\DefWarn#1\xdef#1{fig.~\the\figno}
\writedef{#1\leftbracket fig.\noexpand~\the\figno}%
\figinsert\figin{\centerline{#3}}\medskip\centerline{\vbox{\baselineskip12pt
\advance\hsize by -1truein\noindent\footnotefont{\bf Fig.~\the\figno:} #2}}
\bigskip\endinsert\global\advance\figno by1}
\else
\def\ifig#1#2#3{\xdef#1{fig.~\the\figno}
\writedef{#1\leftbracket fig.\noexpand~\the\figno}%
\global\advance\figno by1}
\fi
\ifamsf
\font\blackboards=msbm7
\font\blackboardss=msbm5
\textfont\black=\blackboard
\scriptfont\black=\blackboards
\scriptscriptfont\black=\blackboardss
\def\Bbb#1{{\fam\black\relax#1}}
\else
\def\Bbb{\bf}
\fi
%
\def\yboxit#1#2{\vbox{\hrule height #1 \hbox{\vrule width #1
\vbox{#2}\vrule width #1 }\hrule height #1 }}
\def\fillbox#1{\hbox to #1{\vbox to #1{\vfil}\hfil}}
\def\ybox{{\lower 1.3pt \yboxit{0.4pt}{\fillbox{8pt}}\hskip-0.2pt}}

\def\({\left(}
\def\){\right)}
\def\comments#1{}

\def\p{\partial}

\def\CE{{\cal E}}

\def\CM{{\cal M}}
\def\CN{{\cal N}}

\def\nl{\hfill\break}

\def\ap{\alpha'}

\def\II{\relax{I\kern-.10em I}}
\def\IIa{{\II}a}

\def\IZ{\relax\ifmmode\mathchoice
{\hbox{\cmss Z\kern-.4em Z}}{\hbox{\cmss Z\kern-.4em Z}}
{\lower.9pt\hbox{\cmsss Z\kern-.4em Z}}
{\lower1.2pt\hbox{\cmsss Z\kern-.4em Z}}\else{\cmss Z\kern-.4em
Z}\fi}
\def\IB{\relax{\rm I\kern-.18em B}}
\def\IC{{\relax\hbox{$\inbar\kern-.3em{\rm C}$}}}
\def\ID{\relax{\rm I\kern-.18em D}}
\def\IE{\relax{\rm I\kern-.18em E}}
\def\IF{\relax{\rm I\kern-.18em F}}
\def\IG{\relax\hbox{$\inbar\kern-.3em{\rm G}$}}
\def\IGa{\relax\hbox{${\rm I}\kern-.18em\Gamma$}}
\def\IH{\relax{\rm I\kern-.18em H}}
\def\II{\relax{\rm I\kern-.18em I}}
\def\IK{\relax{\rm I\kern-.18em K}}
\def\IP{\relax{\rm I\kern-.18em P}}

%

\def\inbar{\,\vrule height1.5ex width.4pt depth0pt}

\def\p{\partial}

\font\cmss=cmss10 \font\cmsss=cmss10 at 7pt
\def\IR{\relax{\rm I\kern-.18em R}}

\ifamsf
\def\IC{\Bbb{C}}
\def\IP{\Bbb{P}}
\def\IR{\Bbb{R}}
\def\IZ{\Bbb{Z}}
\fi

\def\BZ{\IZ}
\def\BP{\IP}
\def\BC{\IC}

\def\lp10{l_P^{10}}
\def\lp11{l_P^{11}}
\def\R11{R_{11}}

\Title{\vbox{\baselineskip12pt\hbox{hep-th/9707214}
\hbox{RU-97-54}
\hbox{CU-TP-850}}}
{\vbox{
\centerline{Metrics on D-brane Orbifolds} }}
\centerline{Michael R. Douglas$^1$ and Brian R. Greene$^2$}
\medskip
\centerline{$^1$Department of Physics and Astronomy}
\centerline{Rutgers University }
\centerline{Piscataway, NJ 08855--0849}
\centerline{\tt mrd@physics.rutgers.edu}
\medskip
\centerline{$^2$Departments of Physics and Mathematics}
\centerline{Columbia University }
\centerline{New York, NY 10025}
\medskip
\bigskip
\noindent
We calculate the metric on the $D$-brane vacuum moduli space for
backgrounds of the form $\BC^3/\Gamma$ for cyclic  groups $\Gamma$.
In the simplest procedure --- starting with a flat ``seed'' metric on
the covering space --- we find that the resulting
$D$-brane metric is not
Ricci-flat. We argue that this is likely to be true of the true 
0-brane metric at weak string coupling.

\Date{June 1997}
%

\lref\bss{T. Banks, N. Seiberg, and E. Silverstein, ``Zero and
One-dimensional Probes with $N{=}8$ Supersymmetry,'' hep-th/9703052.}
\lref\DHVW{L. Dixon, J. A. Harvey, C. Vafa, and E. Witten, ``Strings on
Orbifolds, I, II'' Nucl. Phys. B261 (1985) 678; Nucl. Phys. B274 (1986) 285.}
\lref\itoreid{Y. Ito and M. Reid, ``The McKay Correspondence for Finite
Subgroups of SL(3,\BC),'' in: {\it Higher Dimensional Complex Varieties}\/
(M.~Andreatta et al., eds.), de Gruyter, 1996, p.~221; alg-geom/9411010.}
\lref\reid{M. Reid, ``McKay Correspondence,'' alg-geom/9702016.}
\lref\GanMorSei{O. J. Ganor, D. R. Morrison, and N. Seiberg, ``Branes,
Calabi--Yau Spaces, and Toroidal Compactification of the $N{=}1$
Six-Dimensional $E_8$ Theory,'' Nucl. Phys. B487 (1997) 93;
hep-th/9610251.}
\lref\MPorb{D. R. Morrison and M. R. Plesser, to appear.}
\lref\sag{A. Sagnotti, ``Some Properties of Open-String Theories,''
hep-th/9509080.}
\lref\kron{P. B. Kronheimer, ``The Construction of ALE Spaces as
Hyper-K\"{a}hler Quotients,'' J. Diff. Geom.  29 (1989) 665.}
\lref\infirri{A. V. Sardo Infirri, ``Partial Resolutions of Orbifold
Singularities via Moduli Spaces of HYM-type Bundles,'' alg-geom/9610004.}
\lref\infirritwo{A. V. Sardo Infirri, ``Resolutions of Orbifold Singularities
and Flows on the McKay Quiver,'' alg-geom/9610005.}
\lref\polcai{J.~Polchinski and Y.~Cai, ``Consistency of Open Superstring
Theories,'' Nucl. Phys.  B296 (1988) 91.}
\lref\bwb{M. R. Douglas, ``Branes within Branes,'' hep-th/9512077.}
\lref\dm{M. R. Douglas and G. Moore, ``D-Branes, Quivers, and ALE Instantons,''
hep-th/9603167.}
\lref\dg{M. R. Douglas and B. R. Greene, to appear.}
\lref\JM{C. Johnson and R. Myers, ``Aspects of Type IIB Theory on ALE
Spaces,'' hep-th/9610140.}
\lref\egs{M. R. Douglas, ``Enhanced Gauge Symmetry in M(atrix) Theory,''
hep-th/9612126.}
\lref\polpro{J.~Polchinski, ``Tensors from K3 Orientifolds,''
hep-th/9606165.}
\lref\BFSS{T. Banks, W. Fischler, S. H. Shenker and L. Susskind,
``M Theory as a Matrix Model: A Conjecture,'' hep-th/9610043.}
\lref\ooy{H. Ooguri, Y. Oz and Z. Yin, ``D-Branes on Calabi--Yau Spaces and
Their Mirrors,'' Nucl.Phys. B477 (1996) 407; hep-th/9606112.}
\lref\agm{P. S. Aspinwall, B. R. Greene and D. R. Morrison, ``Calabi--Yau
Moduli Space, Mirror Manifolds and Spacetime Topology Change in
String Theory,'' Nucl. Phys. B416 (1994) 414; hep-th/9309097.}
\lref\rAGMsd{P. S. Aspinwall, B. R. Greene and D. R. Morrison, ``Measuring
Small Distances in $N{=}2$ Sigma Models,''
Nucl. Phys. B420 (1994) 184; hep-th/9311042.}
\lref\aspinwall{P. S. Aspinwall, ``Enhanced Gauge Symmetries and K3
Surfaces,'' Phys. Lett. B357 (1995) 329; hep-th/9507012.}
\lref\dos{M. R. Douglas, H. Ooguri and S. H. Shenker, ``Issues in M(atrix)
Theory Compactification,'' hep-th/9702203.}
\lref\rWP{E. Witten, ``Phases of $N{=}2$ Theories In Two Dimensions,''
Nucl. Phys. B403 (1993) 159; hep-th/9301042.}
\lref\witPT{E. Witten, ``Phase Transitions In M-Theory And F-Theory,''
Nucl. Phys. B471 (1996) 195; hep-th/9603150.}
\lref\fulton{W. Fulton, {\it Introduction to Toric Varieties,}
Princeton University Press, 1993.}
\lref\oda{T. Oda, {\it Convex Bodies and Algebraic Geometry,}
Springer-Verlag, 1988.}
\lref\rGK{B. R. Greene and Y. Kanter, ``Small Volumes in Compactified
String Theory,'' hep-th/9612181.}
\lref\rAG{P. S. Aspinwall and B. R. Greene, ``On the Geometric
Interpretation of $N{=}2$ Superconformal Theories,''
Nucl. Phys. B437 (1995) 205; hep-th/9409110.}
\lref\mp{D. R. Morrison and M. R. Plesser, ``Summing the Instantons:
Quantum Cohomology and Mirror Symmetry in Toric
Varieties,'' Nucl. Phys. B440 (1995) 279; hep-th/9412236.}
\lref\rDelzant{T.~Delzant, ``{H}amiltoniens p\'eriodiques et images convexe de
  l'application moment,'' Bull. Soc. Math. France {\bf 116} (1988) 315.}
\lref\rAudin{M.~Audin, {\it The Topology of Torus Actions on Symplectic
Manifolds}, Birkh\"auser, 1991.}
\lref\rCox{D. A. Cox, ``The Homogeneous Coordinate Ring of a Toric
Variety,'' J. Algebraic Geom. 4 (1995) 17; alg-geom/9210008.}
\lref\gp{E. G. Gimon and J. Polchinski,
``Consistency Conditions for Orientifolds and D-Manifolds,'' Phys. Rev. D54
(1996) 1667; hep-th/9601038.}
\lref\DGM{M. R. Douglas, B.R. Greene and D.R. Morrision,
{\it Orbifold Resolution by D-branes}, hep-th/9704151.}
\lref\dlp{J.~Dai, R.~G.~Leigh and J.~Polchinski, Mod. Phys. Lett. {\bf A4}
(1989) 2073.}
\lref\pol{J.~Polchinski, Phys.~Rev.~Lett.~75 (1995) 4724-4727;
hep-th/9510017.}
\lref\leigh{R. Leigh, Mod. Phys. Lett. {\bf A4}, 2767.}
\lref\witten{E. Witten, ``Small Instantons in String Theory,'' hep-th/9511030.}
\lref\calabi{E. Calabi, Ann. Sci. Ec. Norm. Sup. 12 (1979) 269--294.}
\lref\adhm{M. F. Atiyah, V. G. Drinfeld, N. J. Hitchin, and Y. I. Manin,
Phys. Lett. {\bf A65} (1978) 185.}
\lref\hklr{N. J. Hitchin, A. Karlhede, U. Lindstr\"om and M. Ro{\v c}ek,
Comm. Math. Phys. 108 (1987) 535-589.}
\lref\witadhm{E. Witten, ``Sigma Models and the ADHM Construction of
Instantons,'' J. Geom. Phys. 15 (1995) 215-226.}
\lref{\witstr}{E. Witten, ``Some Comments On String Dynamics,''
hep-th/9507121.}
\lref\kron{P. Kronheimer, ``The construction of ALE spaces as
hyper-kahler quotients,'' J. Diff. Geom. {\bf 28}1989)665.}
\lref\kn{P.B. Kronheimer and H. Nakajima, ``Yang-Mills instantons
on ALE gravitational instantons,'' Math. Ann. {\bf 288} (1990) 263.}
\lref\infirri{A. V. S. Infirri, alg-geom/9610004 and alg-geom/9610005.}
\lref\gimon{E. G. Gimon and J. Polchinski,
``Consistency Conditions for Orientifolds and D-manifolds,'' hep-th/9601038.}
\lref\hst{P. Howe, G.Sierra and P. Townsend, Nucl. Phys. {\bf B221} (1983)
331;\nl
G. Sierra and P. Townsend,
``The gauge invariant N=2 supersymmetric
sigma model with general scalar potential,''  Nucl. Phys. B233 (1984) 289.}
\lref\douglas{M.~R.~Douglas, ``Branes within Branes,'' hep-th/9512077.}
\lref\dm{M. R. Douglas and G. Moore, ``D-Branes, Quivers, and ALE Instantons,''
hep-th/9603167.}
\lref\polgeo{J. Polchinski, }
\lref\IS{Intriligator and Seiberg}
\lref\BFSS{T. Banks, W. Fischler, S. H. Shenker and L. Susskind,
``M Theory As A Matrix Model: A Conjecture,'' hep-th/9610043.}
\lref\DKPS{M. R. Douglas, D. Kabat, P. Pouliot and S. Shenker,
``D-branes and Short Distances in String Theory,'' hep-th/9608024.}
\lref\polwit{J.~Polchinski and E.~Witten, ``Evidence for Heterotic-Type I
String Duality,'' hep-th/9510169.}
\lref\joerev{See S. Chaudhuri, C. Johnson, and J. Polchinski,
``Notes on D-Branes,'' hep-th/9602052, for a recent review.}
\lref\kn{P.B. Kronheimer and H. Nakajima, ``Yang-Mills instantons
on ALE gravitational instantons,'' Math. Ann. {\bf 288} (1990) 263.}
\lref\shenker{S. H. Shenker, ...}
\lref\agm{P.S. Aspinwall, B.R. Greene and D.R. Morrison, Nucl. Phys. B416
(1994)
 414-480; hep-th/9309097.}
\lref\fulton{W. Fulton, {\it Introduction to Toric Varieties,}
Princeton University Press, 1993.}
\lref\oda{T. Oda, {\it Convex Bodies and Algebraic Geometry,}
Springer-Verlag, 1988.}
\lref\fr{W. Fischler and A. Rajaraman, {\it M(atrix) String Theory
on K3}, hep-th/9704123.}
\lref\dko{M. R. Douglas, A. Kato and H. Ooguri, to appear.}
\lref\dvv{R. H. Dijkgraaf, E. Verlinde and H. Verlinde, hep-th/9703030.}
\lref\banksei{T. Banks and N. Seiberg, hep-th/9702187.}
\lref\dcurve{M. R. Douglas, hep-th/9703056.}
\lref\suss{L. Susskind, hep-th/9704080.}
\lref\gk{B.R. Greene and Y. Kanter, ``Small Volumes in Compactified String Theory'',
hep-th/9612181, Nucl. Phys. B, to appear.}
%
%
\newsec{Introduction}

During the last year, developments in nonperturbative string theory have
made it possible to probe the structure of space-time on sub-stringy scales.
In the context of weakly coupled \IIa\ string theory,
this emerges from the dynamics of gauge theory,
the world-volume theories of Dirichlet zero-branes, and
geometrical concepts appear to remain sensible all the way down to
the eleven-dimensional Planck scale $l_p^{11}$ \DKPS.
Short distance geometry is similar to the
long distance geometry probed by fundamental
strings, as one would expect if the two are simply limits of a single
idea of `geometry,' but with interesting differences.

In a recent paper \DGM\
the first steps towards understanding short distance Calabi-Yau geometry were
taken.  Comparing with problems with higher supersymmetry, the most
salient difference is that the metric can depend on an arbitrary function,
the K\"ahler potential.  Thus the questions of what metric is seen by
D$0$-branes and whether it is determined by a local equation of motion can
be studied directly.

In \DGM, a study was made of backgrounds of the form
$\BC^3/\Gamma$ for cyclic subgroups $\Gamma \in SU(3)$,
focusing on comparing the topological
properties probed by D$0$-branes with those probed by
fundamental strings. As has been known for some time, fundamental strings
probe a rich phase structure in Calabi-Yau backgrounds which typically
includes regions which are most directly described in a non-geometrical
abstract
conformal field theory language. These non-geometrical phases emerge from
the linear sigma model analysis of \rWP.
 As pointed out in \agm\ and utilized in
\witPT, if one allows for analytic continuation from geometrical
regions and uses physically motivated parameters, even these non-geometrical
phases appear to have geometrical content in terms of Calabi-Yau spaces
in which part or even the whole manifold has been shrunk to string or
sub-stringy scales.

The newfound power of D-branes  allows us to go beyond
the indirect method of analytically
continued fundamental string conclusions.
In \DGM, it was found that, in
a rather novel manner, the topological properties probed by D$0$-branes in
such backgrounds appear to match the analytically  continued fundamental
string results at short distances.

In this note we go beyond the topological questions considered in \DGM\ and
take a first look at metric properties.
Our approach, although calculationally intensive beyond
the simplest of examples, is a straightforward extension of the
techniques used in \refs{\dm,\JM,\DGM}. Namely, we begin with
D-branes on $\BC^3$, arranged in the regular representation of
the group $\Gamma$. We then truncate to the $\Gamma$-invariant sector
of this D-brane lagrangian.  The result is a gauged sigma model whose
low energy configuration space will be interpreted as sub-stringy space-time.
This prescription can be justified by world-sheet computation at weak
string coupling \dm, while its primary justification at strong
string coupling  is simplicity
and the need for the theory to remain non-singular
in the orbifold limit \egs.

In the case of D$0$-branes and quantum mechanics, quantum effects are
controlled by the dimensionless parameters $g_s/m_i^3$, where the
$m_i$ are the masses of the degrees of freedom being integrated out
to derive the low energy effective theory.  Here the massive degrees
of freedom are strings stretched between image branes, whose mass can
be estimated as $m^2 \sim |\zeta|/\ap^2$ \dos.
Restoring all the $\ap$'s, the condition $g_s/m_i^3 << 1$ translates
into $|\zeta| >> l_{p11}^2$, so for blowup larger than the
eleven-dimensional Planck scale the effective theory can be derived
by classical considerations.

The classical moduli space is a
K\"ahler quotient (the same proposed in \infirri), which
allows direct determination of the quotient space metric. As we shall see,
the explicit calculations are facilitated by methods discussed in
\hklr.  The result depends on the original (or ``seed'') metric of the
gauged sigma model, but even starting with a flat seed metric produces
quite non-trivial quotient metrics.

In  the case of $\BC^2/\Gamma$ for $\Gamma$ lying in $SU(2)$
considered in \refs{\dm, \JM}, the effective theory has eight real
supercharges, and
the quotient construction is of {\it hyperk\"ahler} type \hklr.
This
ensures that the resulting metric is Ricci-flat. In the K\"ahler quotients
considered here, although the metric will be of the correct topological
type (zero first Chern class),
nothing ensures a similar Ricci-flatness condition. In fact,
we shall find in the examples we study, that the resulting metric
is not Ricci-flat.  In the mathematical context,
this was first pointed out by Sardo Infirri \infirri.

The result has different interpretations depending on whether we consider
weak or strong string coupling.  At weak coupling,
we know that fundamental strings do not see a Ricci-flat metric, so
it should  not be surprising that D-branes do not see one either.

In principle the D$0$-brane metric could be computed from world-sheet
computation, in two ways.
One could start with the Calabi-Yau sigma model, introduce general boundary
conditions along the lines of \leigh, and derive the conditions for
an RG fixed point.
One would presumably obtain the Nambu-Born-Infeld action with an
effective metric related to the Ricci-flat metric by $\alpha'$ corrections,
a priori different from those of bulk renormalization.
Alternatively, one could compute the true seed metric of the
gauged sigma model of \DGM\ along the lines of \dm, and then perform
the quotient.  In the orbifold limit, the seed metric is flat, and the
analyticity of conformal perturbation theory implies that the true metric
is an analytic function of the blow-up parameters.
The expectation from both approaches is that, although the precise
metric cannot be determined without additional world-sheet computation,
there is no reason for it to be Ricci-flat.

D$0$-branes are also the fundamental degrees of
strongly coupled \IIa\ theory, and a precise form of this statement
is the M(atrix) theory conjecture \BFSS.
In the regime we study, their dynamics is
governed by eleven-dimensional supergravity, leading inescapably to
the conclusion that they move in a background
Ricci-flat metric. 

Now the original proposal of \BFSS asserts that M theory
is produced only  by taking the large $N$ limit,
and our result offers no contradiction to this statement.
On the other hand, a more recent proposal of Susskind \suss\ asserts
that the finite $N$ theory also has an M theory interpretation, as
the theory with a compactified light-cone dimension.
We see no reason why such compactification would modify the
prediction of supergravity, in which case the model with a flat seed
metric is in contradiction with this proposal.

Now there is no a priori reason to take a flat seed metric in this
context and the simplest response to this result
is to ask whether the model can be
modified to produce a Ricci-flat metric.
The simplest possibility would to be
to define the model with a seed metric tuned to produce
Ricci flatness after the quotient.  Since the quotient metric has zero
first Chern class and the seed metric has more parameters than the
resulting K\"ahler potential, there is no obvious
obstruction to doing this.
However this leaves unanswered
the question of what consistency condition forces
this choice of seed metric.

Although we leave this question for future work, we would
like to point out here that a promising context for answering
it is to consider the theory with more than one D$0$-brane.
Unlike the case of a single D$0$-brane, now it is possible for
massive strings to become light (when component D$0$-branes approach
to distances $d\sim l_{p11}$), and thus quantum effects 
can always become large.
Thus we propose as an interesting goal for future work
the construction of actions which realize the
axioms of \dcurve\ while remaining non-singular in the orbifold
limit, presumably by deforming the seed metric in the $N>1$ version
version of the construction of \DGM.

\newsec{D-branes on Orbifolds and their Metrics}

In this section we briefly review our procedure for discussing
$D$-branes on orbifolds of the form $\BC^3/\Gamma$.
For more details the reader should consult \refs{\dm, \JM, \DGM}.

Our starting point is an ${\cal N} = 4, d = 4$ $U(n)$ supersymmetric gauge
theory where $n = \dim \Gamma$, arising from $D$-brane
`compactification' on
$\BC^3$. We then specify an action
of $\Gamma$ on the fields of the covering theory, and truncate the
${\cal N }= 4$  Lagrangian to the fields transforming trivially under this
action.
The result is a gauged supersymmetric linear
sigma model with a nontrivial superpotential.

Specifically, we take $\Gamma$ to act on the Chan-Paton degrees of freedom
(which ultimately become the D-brane spatial coordinates) in the regular
representation (other possibilities are discussed in \egs\ and \DGM), and on
the $\BC^3$ coordinates $Z^i$ via $Z^i \rightarrow \omega^{a_i} Z^i$ with
$\omega = exp(2 \pi i/n)$ and  $a_1 + a_2 + a_3 \cong 0$ (mod n). The latter
condition ensures that $\Gamma$ lies in $SU(3)$ and hence, from a four
dimensional perspective, we have ${\cal N} = 1$ in the open string sector,
and  ${\cal N} = 2$ in the closed string sector.

In \DGM\ it was found
to be convenient to think of the constraints arising from the superpotential
as if they were $D$-term constraints in an auxiliary linear sigma model.
The latter was shown to yield a vacuum phase structure which meshes
well with that of the analytically continued fundamental string, at  least
as far as topological properties are concerned.

Our present purpose is to understand metric properties. We will do so
by utilizing our  supersymmetric gauged sigma models
 (or ``linear sigma models'') to
produce classical moduli space metrics of this quotient form \hklr.
Specifically, we will determine the metric  in two steps. First, we
 restrict the original flat metric to solutions of $\p W=0$
where $W$ is the superpotential. Second, we
restrict this metric to solutions of the D-flatness conditions
and quotient by the gauge action.  This second step is a symplectic
reduction or ``K\"ahler quotient'' as it is guaranteed to produce a
K\"ahler metric.

We note that if we were to
take the superpotential in the first step to be the one
determined by $\CN=2$, $d=4$ supersymmetry and perform both steps,
the resulting procedure is the hyperk\"ahler quotient.
On general grounds  the hyperk\"ahler quotient is guaranteed to
produce a Ricci-flat metric.
Physically, this would be the relevant construction
if we were considering {\it two} dimensional orbifolds $\BC^2/\Gamma$.
In the three-dimensional case of present interest, though, there  is
no guarantee that the K\"ahler quotient construction yields
a Ricci-flat metric; indeed in the
simplest example, a hypersurface with $c_1=0$ in a $\IP^n$ realized
as the usual quotient of $\BC^{n+1}$  by a $U(1)$ action,
it is not so. We emphasize that this is so in our case even
though we take $\Gamma$
to lie inside of $SU(3)$.
This ensures that we preserve the $c_1 = 0$ condition,
but it does not ensure that the specific metric produced is Ricci-flat.
We shall explicitly see this in what follows.

\newsec{Calculational Procedure}

The two step procedure for determining the
D-brane metric outlined above can be carried out as follows.
For the time being,
we start with the flat K\"ahler potential on the covering space
$$
K_f = \sum_i |z_i|^2,
$$
where the coordinates $z_i$ run over the subset of the original
$3 |\Gamma|$ chiral fields that survive  the orbifold projection.

The first step in the projection is simply to  restrict to the submanifold
by solving the conditions $\del W=0$.
To carry out the second step, we use
the complex reduction procedure of
\hklr.  In physical terms, we write the action
with the coupling to vector superfields $V_i$, and then integrate
them out.  The highest component of $V$ becomes a Lagrange multiplier
for the D-flatness condition.
This action has a complexified gauge invariance which can then be
fixed arbitrarily.

Explicitly,
we write the gauged K\"ahler potential
$$
K = \sum_i ( |z_i|^2 \exp \sum_a t_i^a V_a ) - \sum_a \zeta^a V_a
$$
where $t_i^a$ is the charge of $z_i$ under the $a$'th $U(1)$,
and $\zeta_a$ are the Fayet-Iliopoulos parameters.
We will also let $q_a = \exp V_a$ in the following.

To derive an explicit metric, we choose a gauge slice $X$, with
local coordinates $x^i$.
The gauged K\"ahler potential on the slice
is $K = K(\{x^\mu\},\{q^j\},\{\zeta_k\})$.
We then determine the $q^j$ by solving the equations
\eqn\minimize{
 \partial_j K |_X = 0,
}
and find the metric
$$
g_{\mu \nu} = \partial_{\mu} \partial_{\nu} K|_X.
$$
In these and other expression, $\partial_{\mu}$ and $\partial_j$
are with respect to $x^\mu$ and  $q^j$.
We can also write this as
$$
g_{\mu \nu} = \partial_{\mu} \partial_{\nu} K +
\partial_{\mu} \partial_{j} K \partial_{\nu}q^j +
\partial_{\nu} \partial_{j} K \partial_{\mu}q^j +
\partial_l \partial_j K \partial_{\nu} q^j \partial_{\mu} q^l.
$$
The partials $\p_\mu q^i$ are
determined by considering
$$
{d\over dq^i} \partial_\mu K = 0,
$$
which implies
$$
\partial_{\mu} \partial_i K = - \partial_i \partial_j K \partial_{\mu}q^j
$$
and
$$
\partial_{\mu} q^j = - (\partial_{\mu} \partial_i K) ( \partial_j \partial_i
K)^{-1}
$$
where the latter inverse is in the matrix sense.
Using both of these equations we find
$$
g_{\mu \nu} = \partial_{\mu} \partial_{\nu} K - A^t(B^{-1})^t A
$$
where $A = (A_{i \mu}) = (\partial_i \partial_\mu K)$ and
$B = (B_{ij}) = (\partial_i \partial_j B)$.

This determines the metric explicitly given the ability to solve
the equations \minimize.
In practice, except in the simplest of
examples, this equation must be solved numerically.

We will describe the results in two examples below.
As our interest will be to determine if
the D-brane metrics are Ricci-flat, we first,
as a point of comparison, briefly review relevant work
of Calabi on the construction of Ricci-flat metrics.

\newsec{Ricci flat metrics a la Calabi}

In \calabi, explicit constant curvature K\"ahler metrics are constructed
on non-compact spaces.  One starts with a constant curvature K\"ahler
metric on $\CM$ of complex dimension $n-1$, and writes a metric ansatz
on a line bundle $\CE$ over $\CM$ for which the differential
equation $R_{i\bar j}=c g_{i\bar j}$ can be solved explicitly.
If $\CE$ is the canonical line bundle, the resulting metric will be Ricci-flat.

Let $z^i$ be coordinates on $\CM$ and $K_0(z,\bar z)$ the K\"ahler
potential, satisfying
$$
\det g^{(0)}_{i\bar j} =
\det \p_i\bar\p_{\bar j} K_0 = e^{-l K_0} |(holomorphic)|^2.
$$
Let $w$ be a coordinate on the fiber.
The metric ansatz on $\CE$ is then
\eqn\kansatz{
K = K_0 + u(a(z,\bar z)|w|^2).
}
Since $x\equiv a|w|^2$ is a hermitian form on the fiber, it has an associated
connection $L = \p\log a$ and curvature $S=\bar\p L$.

Metrics derived from the ansatz \kansatz\ can be expressed simply in
terms of the basis of one-forms $dz^i$ and $\grad w = dw + w L$:
$$
\p\bar \p K =
(g^{(0)}_{i\bar j}
  + x u'(x) S_{i\bar k} J^{\bar k}_{\bar j})dz^i d\bar z^{\bar j}
        + (u'(x)+xu''(x)) a |dw + w L|^2
$$
where $J$ is the complex structure.
In this basis, $\det g$ is not hard to compute.

Next, one takes for $a$ the form induced on the canonical line bundle
from the constant curvature metric.
Coordinates can be chosen in which this is just $a = e^{l K_0}$,
and $S_{i\bar j}=l\omega_{i\bar j}$.
The equation $\det g = {\rm const}$ then reduces to
$$
(1 + l x u')^{n-1} (xu'' + u') = {\rm const}.
$$
This has the explicit real solution
$$
u(x) = {n\over l}\(\root n\of{1+cx}-1\) - {1\over l}\sum_{j=1}^{n-1}
(1-\omega^j)\log\({\root n\of{1+cx}-\omega^j\over 1-\omega^j}\)
$$
with $\omega=e^{2\pi i/n}$.
The holomorphic $n$-form is simply
$$
\Omega^{(n)} = dz^1\wedge\ldots\wedge dw.
$$

We now take $\CM=\IP^{n-1}$, and apply this construction to obtain
a Ricci-flat metric on $\CO_{\IP^{n-1}}(-n)$.
For $n=2$ this produces the Eguchi-Hanson metric.  In general Calabi showed
the metric is complete but the ALE nature of the space seems to have
not been explicitly addressed.
Fortunately, since
the asymptotic behavior for large $x$ will be governed by the first term
in $u$, $K \sim x^{1/n} + O(\log x)$ with
$x \sim |w|^2 e^{l K_0}$, it is fairly easy to show that a given example
is ALE.

For $n=3$,
the resulting total space will be $\CO_{\IP^2}(-3)$ which is a blowup
of $\BC^3/\BZ_3$.  The constant curvature metric has
$K_0 = \zeta\log \sum |z_i|^2$ with $l=3$.
The original $\BP^2$ is $w=0$ so $\zeta$ becomes the volume of the
two-cycle.

\newsec{Examples}

In this section, we carry out the K\"ahler reduction to determine
the orbifold and blown-up metric probed by D-branes.

\subsec{Example: the Eguchi-Hanson metric}

As shown in \refs{ \dm, \JM},
the D-brane model realizes Kronheimer's construction as
a hyperk\"ahler quotient of $\BC^4$ by a $U(1)$
action.  Usually this is done in a way which makes the $SU(2)$ acting
on the complex structures manifest, and produces the metric in the
form given by Gibbons and Hawking.  Here we want a single complex
structure manifest, so we do the complex reduction.

We start with four complex fields $b_0$, $b_1$, $\bar b_0$ and $\bar b_1$
with charges $+2$, $-2$, $-2$ and $+2$ respectively.  The original
holomorphic two-form is
\eqn\holotwoform{
\Omega_f = db_0\wedge d\bar b_0 +db_1\wedge d\bar b_1.
}

The gauged K\"ahler potential is
$$
K = e^V (|b_0|^2 + |\bar b_1|^2) + e^{-V} (|b_1|^2 + |\bar b_0|^2 )
        - \zeta_R V
$$
while the constraints $\del W=0$ are
$$
b_0 \bar b_0 - b_1 \bar b_1 = \zeta_C.
$$
The resulting metric is invariant under rotating the parameter $\vec\zeta$,
so we take $\zeta_C = 0$ and relabel $\zeta_R=\zeta$.

Write $q\equiv e^V$, $b_0=z$, $\bar b_0=w$,
set the gauge $\bar b_1=1$ and solve $b_1=zw$.
We then have
$$
K = q(1+|z|^2) + {1\over q}(1+|z|^2)|w|^2 - \zeta\log q.
$$
The condition $\p K/\p q=0$ determines
$$\eqalign{
0 &= q^2 - |w|^2 - {\zeta q\over 1+|z|^2}  \cr
q &= {1\over 2(1+|z|^2)}\(\zeta \pm \sqrt{\zeta^2+4(1+|z|^2)^2|w|^2}\)
}$$
and after substituting back,
$$
K = \pm \sqrt{\zeta^2 + 4(1+|z|^2)^2|w|^2} - \zeta \log q.
$$
Notice that we have made a fortuitous choice of coordinates:
the combination $(1+|z|^2)^2|w|^2$ is the $x$ of Calabi's construction,
and we get the metric exactly in his form.
(The prefactor $\zeta$ is controlling the overall scale and could be
put inside by change of coordinates.)
The simple way to find this is to look for a  choice which turns
\holotwoform\ into $dz\wedge dw$.

\subsec{$\BC^3/\BZ_3$}

Now, as in \DGM,
 the model starts with nine complex fields $X^i$, $Y^i$ and $Z^i$;
there are two $U(1)$ actions which we associate with vector superfields
$U$ and $V$.  As before, let $U = \log p$ and $V = -\log q$, then
the gauged K\"ahler form is
\eqn\kahlerthree{\eqalign{
K =
&{1\over pq}(|X^0|^2 + |Y^0|^2 + |Z^0|^2)
+ q(|X^1|^2 + |Y^1|^2 + |Z^1|^2)
+ p(|X^2|^2 + |Y^2|^2 + |Z^2|^2) \cr
&- \zeta_1 \log p - \zeta_2 \log q.
}}
There are nine  constraints $X^i Y^j = X^j Y^i$,  $X^i Z^j = X^j Z^i$
and $Y^i Z^j = Y^j Z^i$
 of which four are independent, leaving us with a five dimensional
solution space.
By
using complexified gauge transformations we can go to the
three-dimensional slice $Z_1=Z_2=1$.
The holomorphic three-form is
\eqn\holothreeform{\eqalign{
\Omega^{(3)} &= \sum_i dX^i \wedge dY^{i+1} \wedge dZ^{i+2} \cr
&\rightarrow dX^1 \wedge dY^2 \wedge dZ^0
}}
which will be non-singular assuming these are single-valued coordinates.
The constraints are solved by
 $Y^2=Y^1$, $X^2=X^1$, $Y^0=Y^1 Z^0$, $X^0=X^1 Z^0$.

Comparison of three-forms and the symmetry of the situation motivate
the identification
\eqn\coords{z_1 = X^1 \qquad z_2 = Y^1 \qquad w = Z^0,}
after which \kahlerthree\ becomes
$$
K = \(1+|z_1|^2+|z_2|^2\)\(p+q+{|w|^2\over pq}\)
- \zeta_1\log p
- \zeta_2\log q.
$$
$p$ and $q$ can again be eliminated explicitly.
Let $A=1+|z_1|^2+|z_2|^2$, $B=A|w|^2$, then
$$
K = A (p+q) + {B \over pq} - \zeta_1 \log p - \zeta_2 \log q
$$
and $\p K=0$ implies
$$\eqalign{
0 &= p^2 q A - pq\zeta_1 - B \cr
0 &= p q^2 A - pq\zeta_2 - B
}$$
which combine to
$$\eqalign{
(p-q)A &= \zeta_1 - \zeta_2 \cr
(p-q)B &= pq(q\zeta_1 - p\zeta_2).
}$$
Let $u=p-q$ and $v=A(q\zeta_1-p\zeta_2)/(\zeta_1-\zeta_2)$,
then
$$\eqalign{
u &= (\zeta_1-\zeta_2)/A \cr
p &= (v + \zeta_1)/A \cr
q &= (v + \zeta_2)/A \cr
x &= {A^2 B} = v(v+{\zeta_1})(v+{\zeta_2}) \cr
K &= \(2 + {1\over \zeta_1-\zeta_2}\)v
 - \zeta_1\log(v + \zeta_1)
 - \zeta_2\log(v + \zeta_2)
 + (\zeta_1 + \zeta_2)\log A
}$$

Notice that this expression shows a strong resemblance to Calabi's expression.
In particular, if we take
$x = v (v + \zeta_1)(v + \zeta_2)$ with $\zeta_1 = 1 -
\omega$ and $\zeta_2 = 1 -\omega^2$ the result is exactly that of
Calabi.
However, the Fayet-Iliopoulos parameters
are physically (and mathematically)
dictated to be {\it real}. Therefore, the choice we have
taken in order to make contact with Calabi's work is not physical;
at best it is an analytic continuation from the physical domain. For
real values of $\zeta_1, \zeta_2$ we find that the resulting metric is
{\it not} Ricci-flat. Thus, although we have come tantalizingly close
to a Ricci-flat metric, in reality the quotient construction will not
produce one except in the trivial case $\zeta=0$.

The non-Ricci flatness of this metric was first found
in \infirri\ using a less explicit argument.  The construction
presented here not only gives the explicit form of the quotient
metric but indeed allows writing explicit one-parameter families
of metrics interpolating between the Ricci-flat and quotient
metrics, by allowing the parameters $\zeta$ to move off into the
complex plane.
The physical D$0$-brane
metrics form such a family, interpolating between Ricci-flat for
$\zeta>>\ap$ and a true quotient metric for $\zeta<<\ap$,
and one is led to wonder if they can be described in this way.

It is also interesting to note that while the Ricci-flat metrics form
a family with one real parameter (the volume of the $S^2$), the
these non-Ricci flat metrics naturally depend on two real parameters,
the $\zeta_i$.
The volume of the $S^2$ corresponds to one linear combination (which
one depends on the signs of the $\zeta$'s as described in \DGM),
but the full metric depends on both.  The other parameter becomes
$\int B$ in the large blowup limit, and in this sense we see that
the sub-stringy geometry explicitly depends on moduli which were
non-metric moduli at large volume.

\subsec{$\BC^3/\BZ_n, n > 3$}

For $n > 3$, the analogous equations are easy to write down but
do not appear to admit analytic solutions for general $\zeta$.
However, they are certainly amenable to numerical evaluation.
To assess
whether or not the metric is Ricci-flat, it is simplest to
compute $\det g$ and see whether or not it is constant over
the manifold.  In the present circumstances, constant  $\det g$
is equivalent to Ricci-flatness.  We have done this for $n = 5$ and $n = 7$
for various choices of the Fayet-Iliopoulos parameters.  We find that
the metric, once again, is {\it not} Ricci-flat.

\newsec{Conclusions}

For weak string coupling and sub-stringy blowup parameters,
D$0$-branes propagating on an ALE space of three complex
dimensions see a metric which is calculable in principle by a
combination of world-sheet techniques and the quotient construction
described here.  It depends on the full complexified K\"ahler class,
not just the real part.
There is no reason to expect the result to be Ricci-flat.

We gave explicit forms for the Ricci-flat metric on $\BC^3/\BZ_3$
and for the quotient metric obtained by starting with a flat
seed metric.  Although we found that the quotient construction
could indeed produce a Ricci-flat metric in this case, this was
obtained by a rather formal procedure of ``complex reduction with
complex Fayet-Iliopoulos parameters'' for which we did not find
a satisfactory physical (or mathematical) interpretation.  (It would
be interesting if there were one.)  We also found no evidence that
similar adaptation of the quotient construction could produce Ricci
flat metrics in general, and suspect that it arose from particular
simplicities of the $\BZ_3$ case.  It does provide a nice family
of metrics interpolating between Ricci-flat and quotient metrics.

In the M theory limit, we believe it is possible to postulate
a seed metric which reproduces the expected physical result,
that after quotienting a D$0$-brane sees a Ricci-flat metric.
We furthermore pointed out that the same construction applies
to bound states of $N>1$ D$0$-branes, and that quantum effects
can become large in this case, possibly leading to new consistency
conditions.

We close by listing further
interesting questions to address in the present framework:

\item{1.}
What is the geometric interpretation of the dependence of the
metric on the complexified K\"ahler class?
We know that in quantum geometry the effective volumes of two-cycles
and four-cycles can be controlled independently
\gk; this should
translate into statements about the masses of wrapped branes outlined
in \DGM, but can we relate these brane masses to the explicit metric?
Can we postulate a $B$-field on the two-cycle which reproduces quantum
volumes as proposed there?

\item{2.}
We argued that on grounds of genericity there exists a seed metric
for which the quotient metric will be Ricci-flat.
Is it true that this seed metric
will be the flat metric plus corrections analytic in $\zeta$, as we
might expect to get either from conformal field theory considerations
or from the background considerations in \dos ?  This would allow
corrections expressed as a double series in $\zeta/\ap$ and $\zeta/z^2$.
An even stronger condition which might emerge from deeper study of
the world-sheet computation is that the corrections to the seed
metric should be real analytic in the coordinates as well.  This would
preclude $\zeta/z^2$ corrections and
imply that in the limit $|\zeta|<<\ap$ the D$0$-brane metric reduces
to the quotient metric computed here.

\item{3.}
Is there any simple equation which the true weak coupling D$0$-brane metric
should satisfy (e.g. a $\beta$-function equation with all $\ap$
corrections included).  Because this metric is physically observable,
this question is better motivated than for the analogous fundamental
string metric.
One possibility (suggested by G. Tian) is to try
$0 = R_{ij} + \delta S/\delta g_{ij}$ for $S$ some generating
functional for an anomaly of a six-dimensional Dirac operator.
$S$ would be an eight-form constructed from curvatures, which fits
with the known three-loop sigma model correction, but we have no
real justification for this idea at present.

\medskip
We would like to thank O. Aharony, S. Kachru, D. Morrison, E. Silverstein,
L. Susskind and  G. Tian
for useful conversations.

\bigskip

\listrefs
\end